# Magnetic Charge Search for the BELLE II Detector*

M. K. Sullivan, D. Fryberger

*SLAC National Accelerator Laboratory*
*2575 Sand Hill Rd. MS-39 and MS-51, Menlo Park, CA, USA 94025*
*sullivan@slac.stanford.edu, fryberger@slac.stanford.edu,*

The introduction of magnetic charge into Maxwell's equations has led to an extensive search for magnetically charged particles (magnetic monopoles). A particle model developed by one of us (DF) adds an additional feature to Maxwell's symmetric equations in that the stable magnetic monopole should have the same charge strength as the electron. We have not found any experiments in high-energy physics that have explicitly ruled out this possibility. However, the few experiments at colliders that had no magnetic field might have observed a signal for these 1$e$ strength magnetic monopoles as an unexpected enhancement in the $\mu^+\mu^-$ production rate. The absence of any such observation leads us to set a tentative lower mass limit for these unit charge magnetic monopoles at 4.5-5 GeV/$c^2$. Using a MC generator for magnetic charge and tracking these events through a simplified model of the BELLE II detector, we have found that the central drift chamber of BELLE II has a remarkably high efficiency for triggering on magnetically charged tracks. We suggest that the BELLE II collaboration perform a specific search for stable magnetically charged particles having a field strength of 1$e$ when they run for the first time with colliding beams in 2018. This would be the first time anyone has specifically looked for such a particle.

* Supported by the Department of Energy Contracts DE-AC02-76SF-00515 and DE-AC05-84ER40150

## I. INTRODUCTION

The search for magnetic charge has been going on since Dirac first proposed that the existence of magnetic charge could explain the quantization of electric charge in 1931 [1]. Nearly all searches have been for a highly charged magnetic monopole of magnetic strength ~68.5$e$ or higher, which is commonly called a Dirac monopole. We propose here that magnetic monopole searches should also look for magnetic charge as low as 1$e$ field strength. We have not found any publications that describe an experiment that specifically looked for magnetic charge at the 1$e$ strength level. CESR at Cornell did look for magnetically charged particles down to about the 2$e$ level and did not see anything [2]. Non-magnetic detectors at $e^+e^-$ colliders (such as the Crystal Ball [3] at PETRA and the Free Quark Search [4] at PEP-I) could have seen the production of magnetically charged 1$e$ strength particles by observing what would have appeared to be an over-abundance of muons. The model described in part below predicts the existence of a stable magnetic monopole of 1*e* field strength and also predicts that such a particle should produce an ionization signal in standard electric matter that is twice the ionization signal of an ordinary muon [5]. Another difference between muons and these magnetic monopoles is that the monopole ionization signal is generally flat [6]. In addition, the production rate of these stable 1*e* magnetic monopoles (we call them magneticons) has a $\beta^3$ threshold rate dependence but should otherwise match the production rate for continuum muon pair production in $e^+e^-$ colliders or in *pp* colliders using the Drell-Yan process [5]. These differences in the ionization signal and in the production rate make it difficult to interpret results from previous experiments in terms of whether or not a particular search might have seen this proposed magnetic monopole. We have studied previous experimental data and have tentatively concluded that a minimum mass for the magneticon is about 4.5 GeV/$c^2$. Earlier experiments (e.g. Bubble chambers), at lower production masses tended to be more inclusive and we feel that a lower mass magneticon might have been seen in some of these experiments. However, it is not at all clear that at even lower masses such a particle would have been detected unless a specific attempt was made to look for the unique characteristics of such an event (i.e. magnetically charged track trajectories). The lower limit of 4.5 GeV/$c^2$ is primarily based on the absence of an observed signal (no enhancement in the $\mu^+\mu^-$ cross-section) in the Crystal Ball data while it was running at PETRA with an $E_{cm}$ of 10.58 GeV or on the $\Upsilon(4S)$ [7].

We first describe some of the details of the theory that predicts the existence of these magneticons and call into question some of the assumptions Dirac and others used to predict the magnetic monopole charge strength to be 68.5*e*. We then use a very simple simulator of the BELLE II detector to show how 1*e* magnetically charged particles would look in the Central Drift Chamber (CDC). We find that the BELLE II CDC is remarkably efficient in detecting magnetically charged particles, and if specific triggering, track finding and track fitting algorithms can be constructed, the BELLE II collaboration should be able to conduct, for the first time, a dedicated search for these unit charge magnetic monopoles.

## II.  TOTALLY SYMMETRIC MAXWELL'S EQUATIONS

*Maxwell's Equations with magnetic charges and currents*

We employ the set of Maxwell's equations that are totally symmetric. We use Gaussian units where the magnetic field has units of gauss and the electric field has units of statvolts/cm.

$$\vec{\nabla} \cdot \vec{D} = 4\pi \rho_e, \qquad \vec{\nabla} \times \vec{H} - \frac{1}{c}\frac{\partial \vec{D}}{\partial t} = \frac{4\pi}{c} \vec{j_e}, \qquad (1), (2)$$

and

$$\vec{\nabla} \cdot \vec{B} = 4\pi \rho_m, \qquad -\vec{\nabla} \times \vec{E} - \frac{1}{c}\frac{\partial \vec{B}}{\partial t} = \frac{4\pi}{c} \vec{j_m} \qquad (3), (4)$$

Note that the induced electric field from moving magnetic charge has a minus sign with respect to induced magnetic fields from moving electric charge (the vector curl term in Eqs. 2 and 4). In this analysis, the electric and magnetic charge are both scaler quantities and consistency is maintained by the use of two vector potentials [8].

*Dirac charge quantization from magnetic monopoles*

Dirac observed that the existence of a magnetic monopole implies the quantization of electric charge [9]. The argument was illustrated by Saha [10] using a point electric and a point magnetic particle separated by distance *d* (see Fig. 1). An electric point source and a magnetic point source separated by *d* produce a nonzero Poynting vector value for all points in space (except along the axis defined by *d*). The Poynting vector rotates about the axis defined by the separation. If we integrate the Poynting vector over all space we get a value for the angular momentum around this axis. The integral depends only on the strength of the sources, namely, $\frac{eg}{c}$. The distance *d* falls out of the integral. If we quantize this total angular momentum in units of $\frac{\hbar}{2}$, we have $\frac{eg}{c} = \frac{n\hbar}{2}$ where *n* is an integer and, for n=1 we have a smallest value of the angular momentum. Thus we have quantized the product *eg*. It is interesting to note here that electric and magnetic charge quantization by this method is therefore a direct consequence of angular momentum quantization. Now we also know that $\alpha = \frac{e^2}{\hbar c} \cong \frac{1}{137}$. We therefore can solve for *g* as a function of *e* and we get $g \cong 68.5e$, which in turn leads to $\alpha_m = \frac{g^2}{\hbar c} \cong \frac{(68.5e)^2}{\hbar c} \cong \frac{(68.5)^2}{137} \cong 34$. We find that the magnetic charge quantized in this manner is much stronger than the quantized electric charge.

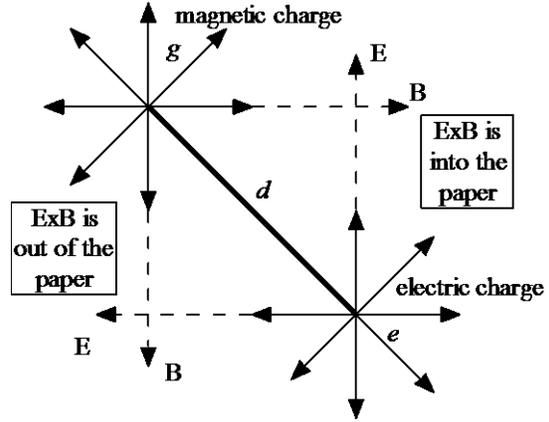

Figure 1. Diagram of how a magnetic monopole and an electric monopole separated by distance *d* produce angular momentum through the Poynting vector.

There are at least two distinct assumptions that go into this calculation by Dirac and Saha. One is that the sources of electric and magnetic field are *point* sources. The sources must be points in order for the distance *d* between the sources to drop out of the integral. This assumption runs counter to the present notion that the fundamental particles have sub-structure. Even the super-string theory has finite-sized (although close to Planck length) strings [11]. One other assumption that Dirac made is that there is only one field potential for both the magnetic and electric charges. This assumption does not allow for a fully symmetric set of Maxwell's equations and is the main reason Dirac had to introduce a line singularity into the electromagnetic potential. Fryberger's theory assumes there is a distinct magnetic potential separate from the standard electric potential [12]. This implies a separate magnetic photon from the known electric photon. In quantum mechanics, then, the two photons differ by a $\gamma_5$ matrix which means they have opposite parities.

*A second solution to Maxwell's symmetric equations*

Noether's Theorem [13] states that a symmetry in the Lagrangian implies a conserved quantity and in the case of Maxwell's equations Rainich [14], Cabibbo and Ferrari [15], and Han and Biedenharn [16] all found an angle of rotation (called *dyality* angle by Han and Biedenharn) about the plane defined by the electric and magnetic charge or fields. For instance, the magnetic charge (field) would be the X axis and the electric charge (field) would be the Y axis. Symmetric Maxwell's equations are invariant under this dyality angle rotation. We also have a conserved quantity, namely, an angular momentum associated with this rotation angle or dyangular momentum.

Another solution to Maxwell's totally symmetric equations has also been found [17]. This solution includes the source terms and is a static (time independent) solution. This new solution contrasts with the photon, which is a source-free propagating solution. The name *vorton* has been coined for this second solution to Maxwell's equations. The vorton has a static Coulomb field arising from a nonsingular spherically symmetric charge distribution. It also has two dipole fields generated by a double rotation of the charge distribution. One rotation is about the Z axis (phi rotation) and another is a toroidal rotation about a ring of radius *a* located in the XY plane. The radius *a* specifies the size or scale of the vorton. The two rotations are synchronized such that one rotation about the Z axis takes place in the same increment of time as one rotation about the toroidal axis given by the circle at radius *a*. The double rotation gives the vorton a non-zero Hopf [18] or topological charge which implies the vorton is topologically stable (Hopf charge is conserved). Like Maxwell's equations, the vorton is also invariant under a dyality rotation. It is essentially a "knot" of electro-magnetic field energy. The rest mass of the vorton is defined as the integral of the field energies, and it has been shown that the monopole field energy coming from the spherically symmetric charge distribution equals the sum of the two separately equal dipole field energies. It has also been shown that the charge of a vorton to be 25.83*e* based on the above constraints and using a semi-classical quantization of the angular momentum of the two rotations [19]. This large natural charge of the vorton inspires the following discussion.

*Vorton pairs as a sub-structure of elementary particles*

Using the large inherent charge of the vorton and the extra degree of freedom from the dyality angle one can construct the known set of elementary particles as pairs of bound vortons. In the case of the electron we would have two vortons with dyality angles that are nearly all magnetic charge but with opposite signs and the dyality angles are slightly rotated toward the negative electric direction by ½*e*. The magnetically bound vorton system then has no net magnetic monopole field (the magnetic charges are equal and opposite) and a residual electric charge with the field strength of the electron. In a similar manner, one can construct the remaining known elementary particles [20, 21].

By the same token and because of the full symmetry of this theory, one can expect that there is a magnetic counterpart to all of the elementary electric particles [22]. Thus, there should exist a lowest mass, *stable* magnetic counterpart to the electron (as previously mentioned we refer to this particle as the magneticon) as well as magnetic versions of all the other fundamental electric particles. These new magnetic particles are also bound vorton pairs with dyality angles very nearly along the electric axis and with a small magnetic charge left over after the strong and opposite electrically charged vortons bind together. As we stated earlier, we believe the mass of the lowest magnetic state (the magneticon) is at or above about 4.5 GeV/$c^2$. We have searched the literature for other experiments at higher center-of-mass energies and have not found any experiment at $e^+e^-$ colliders or $pp$ colliders or beam dumps that has explicitly ruled out the existence of a stable magnetic monopole with magnetic charge of 1$e$. In addition, we have also looked at many non-accelerator monopole searches and again have not found any result that rules out the existence of the magneticon. The search included cosmic ray experiments.

## III. A SEARCH FOR THE MAGNETICON IN BELLE II

### The CDC of the BELLE II detector

We suggest that BELLE II perform a search for the magneticon in the first physics run of 2018. The present schedule of the superKEKB and BELLE II is to have the Phase II run in the spring and early summer of 2018 [23]. This run will have the detector and the final focus cryostats installed and online, but the vertex tracking systems (PXD and SVD) will not be installed. Instead special background detectors and small sections of the PXD and SVD detectors are planned to be installed. However, the full drift chamber (CDC) is already installed in the detector and has taken cosmic-ray data. We think the CDC with a stand-alone trigger can look for magnetically charged particles. The CDC has an inner radius section of 8 layers of drift cells that have a smaller cell size than the rest of the CDC [24]. The small-cell section starts at a radius of 16 cm and goes out to a radius of 24 cm. The Z extent of this part of the chamber goes in the forward direction down to 17° from the detector central axis and in the backward direction (–Z) 30° from the central axis. We have constructed a detector simulator that roughly models the magnetic field of BELLE II and also contains the volumes of the small-cell region of the CDC as well as the central part of the CDC. We understand that the small-cell region has a separate trigger for cosmic rays [25]. For a trigger, we have asked that all 8 layers of the small-cell region fire on a track. For simplification, we assume that the layers are 100% efficient.

We understand [26] that if the accelerator tune-up goes reasonably well during the Phase II run then the BELLE II team will ask to run at the $\Upsilon$(6S) which is at 11.02 $E_{cm}$ and that they hope to collect about 20 fb$^{-1}$ of integrated luminosity. With our MC generator and detector simulator, we have made several runs using various mass values for the magneticon. It turns out that if a small-cell CDC trigger, as described above, can be employed then the CDC is remarkably efficient in triggering on magnetically charged particles. Another important fact about this super B-Factory is the boost given to the $E_{cm}$ system by the asymmetric beam energies (7×4 GeV at the $\Upsilon$(4S) or 7.29×4.17 GeV at the $\Upsilon$(6S)). We will elaborate on this below.

As stated earlier, the production rate for the magneticons is expected to be the same as the production rate for muon pairs with the exception that the magneticon rate will be reduced by a $\beta^3$ factor. This is shown in equation (5) below for magneticon anti-magneticon production.

$$\frac{d\sigma_{m\bar{m}}}{d\Omega} = \frac{\alpha\alpha_m (\hbar c)^2 \beta^3}{4s}\left(1+\cos^2\theta\right) \qquad (5)$$

Here $\alpha_m = \alpha \cong 1/137$ by dyality symmetry; $\beta$ is the usual relativistic factor of the produced magneticons in the center of mass (CM) frame; and $s$ is the CM energy squared. The large $\mu^+\mu^-$ production cross-section (see below) and the hoped for integrated luminosity during Phase II should produce a very good rate of detectable magneticons even up to very near the kinematic limit given by the $E_{cm}$. This is despite the rate loss due to the $\beta^3$ factor. Keep in mind that the $\beta^3$ reduction factor applies to the $\beta$ in the $E_{cm}$ or production reference frame. The $\beta$s in the lab frame will be significantly higher because of the superKEKB boost. Therefore we do not anticipate any difficulties concerning trigger thresholds with respect to the CDC trigger even though the flat ionization of the magneticon is smaller than the ionization of a standard electric particle at low momenta.

*Predicted results for a magneticon search in the BELLE II CDC*

Table 1 shows our predicted rates for magneticon production and our estimated efficiency for a CDC trigger as described above. We use 10 fb$^{-1}$ as our integrated luminosity. We calculate the $\mu^+\mu^-$ production cross-section at 11.02 GeV $E_{cm}$ to be 715.2 pb. Therefore 10 fb$^{-1}$ should generate 7152000 muon events. The magneticon production rate is predicted to be the $\mu^+\mu^-$ rate reduced by the factor of $\beta^3$ mentioned above. As one can see in the table the small-cell part of the CDC is quite efficient at detecting the magneticon nearly up to the kinematic limit or 5.51 GeV (half of the total available energy) when only one of two tracks is demanded. Following the table are pictures of the expected tracks for a magnetic monopole of charge 1$e$ for various particle masses. Note the significant curvature of these tracks in the RZ view, which is a distinctive feature of magnetic charge in a magnetic solenoidal field. In order to separate and emphasize the very different track trajectories of the north and south monopoles in the detector magnetic field we have plotted the north monopole tracks in the +R direction of the plot and the south monopole tracks in the –R direction. As one can see, the north monopoles are further accelerated in the +Z direction whereas the south monopoles, while initially boosted in the +Z direction, actually start to turn around as they are accelerated in the –Z direction. This is where the boost of the $E_{cm}$ greatly increases the detection efficiency of the CDC. The south monopoles have an excellent probability of traversing the small-cell volume of the CDC even when the magneticon mass is nearly at the kinematic limit. This is shown below in Figures 9-12 as the magneticon mass approaches the kinematic limit of 5.51 GeV/$c^2$.

Figure 8 shows magneticon tracks for a mass value of 5.4 GeV/$c^2$. We can see that many of the north tracks go through the small-cell section of the CDC. In addition, many of the south tracks traverse the entire CDC. However, in Figure 9, where the magneticon mass is increased to 5.5 GeV/$c^2$, we can see that no north magneticon track (the +R tracks) now reaches the CDC. Requiring two tracks in the trigger will limit the mass range search to 5.47 GeV/$c^2$ (see Table 1). On the other hand, if single track events are accepted then the upper limit of the mass range extends to 5.507 GeV/$c^2$ in which each magneticon has only 3 MeV of kinetic energy in the CM reference frame.

Table 1. Magneticon detection efficiency using the CDC small-cell region with the trigger described in the text. The CM energy is 11.02 GeV. We also include CDC geometric efficiency and event totals for one or two track events as well as exactly two track events. We use 10 fb$^{-1}$ for an integrated luminosity in order to obtain the number of events shown in the last two columns.

| Magneticon | | | CDC small cell geom. eff. | | Triggered events (10 fb$^{-1}$) | |
|---|---|---|---|---|---|---|
| Mass (GeV/$c^2$) | $E_{cm}\beta$ | $E_{cm}\beta^3$ | One or two tracks | Two tracks | One or two tracks | Two tracks |
| 4.5 | 0.577 | 0.192170 | 0.95931 | 0.8562 | 1318475.2 | 1176760.9 |
| 5 | 0.420 | 0.074183 | 0.9666 | 0.8229 | 512838.3 | 436597.0 |
| 5.1 | 0.379 | 0.054238 | 0.9672 | 0.8055 | 375184.3 | 312459.7 |
| 5.2 | 0.331 | 0.036163 | 0.9725 | 0.8695 | 251524.5 | 224884.9 |
| 5.3 | 0.273 | 0.020447 | 0.9747 | 0.8338 | 142538.5 | 121933.5 |
| 5.4 | 0.200 | 0.007978 | 0.9664 | 0.7372 | 55144.4 | 42065.9 |
| 5.45 | 0.147 | 0.003207 | 0.9583 | 0.5542 | 21980.6 | 12711.7 |
| 5.47 | 0.120 | 0.001740 | 0.9430 | 0.3156 | 11736.0 | 3927.8 |
| 5.48 | 0.104 | 0.001132 | 0.9336 | 0 | 7556.4 | 0.0 |
| 5.5 | 0.060 | 0.000218 | 0.8111 | 0 | 1266.8 | 0.0 |
| 5.505 | 0.043 | 0.000077 | 0.6497 | 0 | 359.0 | 0.0 |
| 5.507 | 0.033 | 0.000036 | 0.3866 | 0 | 99.3 | 0.0 |

Figures 4 and 5 below show tracks made by a magneticon with a 4.5 GeV/$c^2$ mass. The end view tracks in Fig. 5 are perfectly straight and hence an explicit straight-line track finder and fitter in this view should select candidates for magnetic charge. A good straight line fit of course implies infinite momentum tracks. Straight line tracks in the end-view are a unique signal of magnetic charge in a solenoidal field.

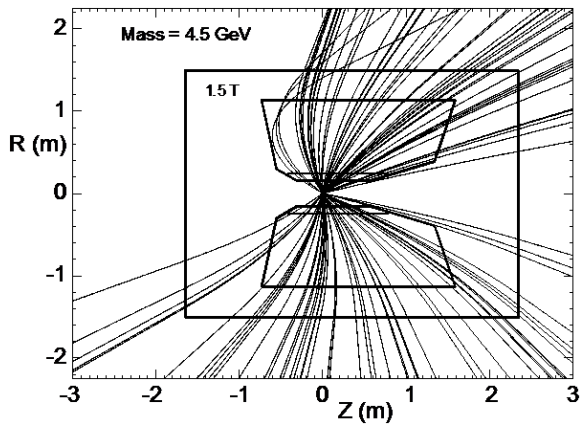

Figure 4. An RZ plot of 4.5 GeV/$c^2$ mass magneticon events that fire the trigger in the small cell CDC. In order to more clearly distinguish the monopole track characteristics we plot the north monopole particles in the +R direction and the south monopole particles in the –R direction.

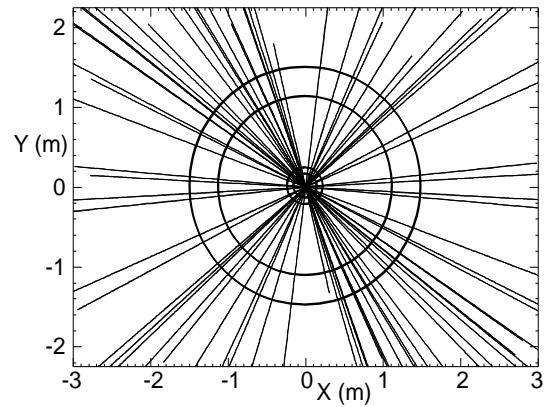

Figure 5. Display of the same events as in Fig. 4 but from the end view of the detector (R,φ). The tracks are perfectly straight in this view.

*Z information in the CDC*

In order to positively identify these events as magnetic, some Z information would be helpful since a magnetic track in the RZ view is also unique. We understand [27] that there is Z information but only in the main CDC where stereo wires have been installed. This means that tracks that go *only* through the small-cell CDC will have no Z information (see Figs. 10 and 12). This happens when the mass approaches the kinematic limit. Figures 6 and 7 show magnetic tracks for magneticon masses of 5.0 and 5.2 GeV/$c^2$ respectively. Here we see that most of the events have tracks that traverse all of the main CDC so these events should have excellent Z information, and the unique RZ tracks for magnetic charge should be detectable. We even see in Figs. 8 and 9, where the magneticon mass is 5.4 and 5.48 GeV/$c^2$, respectively, that at least most of the south magnetic tracks (the −R tracks) still traverse much of the main CDC, and hence we should still have valid Z information for magneticon masses up to 5.48 GeV/$c^2$.

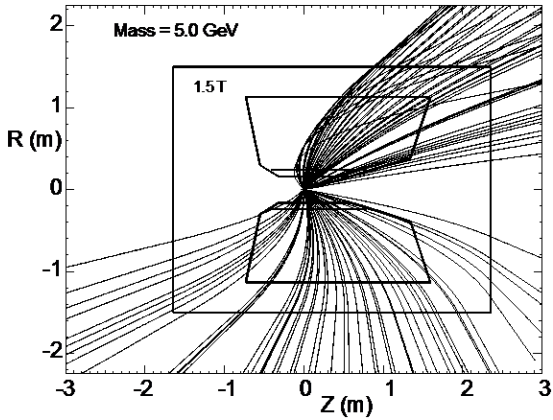

Figure 6. Tracks from a 5.0 GeV/$c^2$ magneticon mass. At least one of the two tracks trigger the small cell CDC.

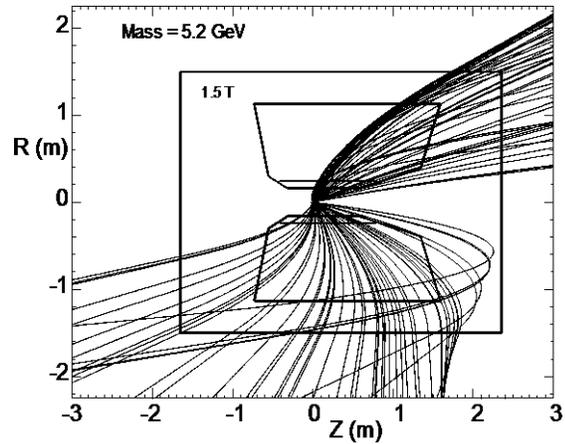

Figure 7. Triggered tracks from a 5.2 GeV/$c^2$ magneticon mass.

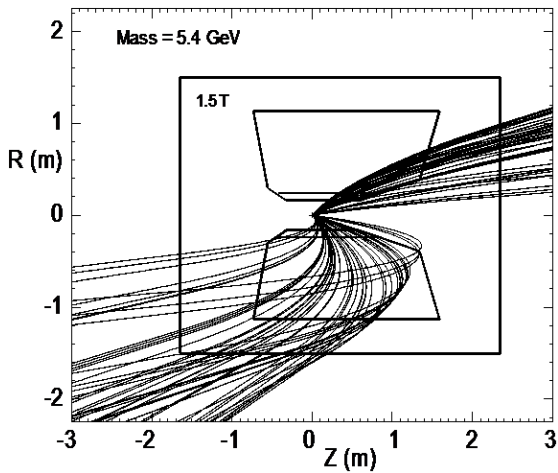

Figure 8. Triggered tracks from a 5.4 GeV/$c^2$ magneticon mass.

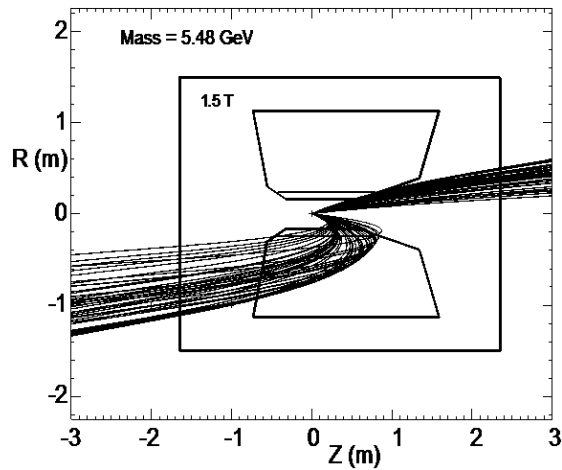

Figure 9. Plot of triggered tracks for a magneticon mass of 5.48 GeV/$c^2$. These events have only one track seen in the small cell CDC. The north magneticons now miss the small cell CDC.

With a magneticon mass of 5.5 GeV/$c^2$ (see Fig. 10) the remaining south tracks (the north tracks go down the beam pipe) are compressed to the small radius part of the main CDC volume. The first 6 layers of wires in the main CDC region are stereo wires so there will be some Z information. However, a detailed detector MC study is likely needed to see if magnetic tracks that travel only in the small-cell region and the very inner part of the main CDC region have enough Z information to unambiguously identify the track as magnetic.

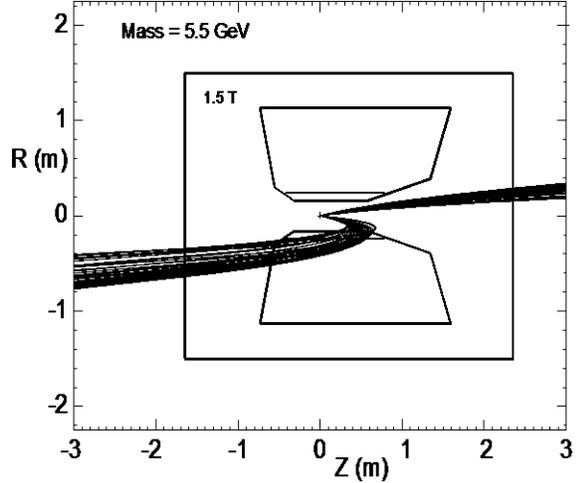

Figure 10. Triggered tracks from a 5.5 GeV/$c^2$ magneticon mass.

At magneticon masses above 5.5 GeV/$c^2$ the south magneticon tracks will only traverse the small-cell CDC and we will have no Z information for these events. See Figs. 11 and 12. However, there is still a unique signature for a magnetic track because the track in the small-cell CDC would appear as an infinite momentum track (i.e. a straight line in the end view), and there would be no other information in the rest of the detector. If such magnetic charge candidates are found, we may have to wait for the VXD installation where there will be more Z information before concluding that magnetic monopoles have been detected.

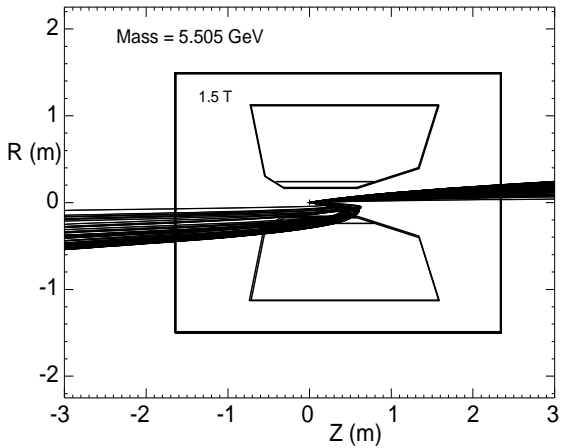

Figure 11. Tracks from a 5.505 GeV/$c^2$ magneticon mass particle. These tracks have only 5 MeV of kinetic energy in the $E_{cm}$ frame.

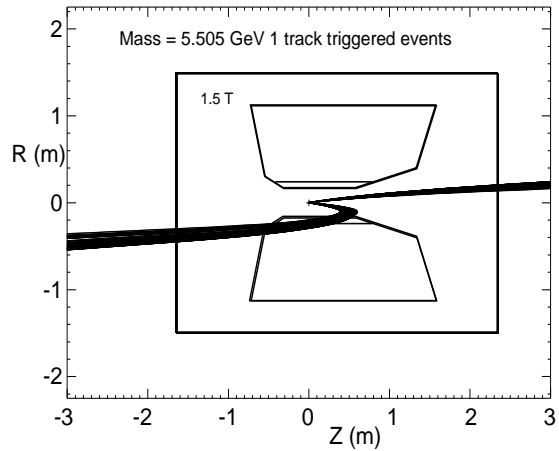

Figure 12. Tracks from a 5.505 GeV/$c^2$ magneticon mass. This is the subset of tracks in Fig. 11 that satisfy the trigger in the small cell CDC.

Nevertheless, even without the VXD, the CDC should be able to positively identify 1$e$ strength magnetic monopoles with masses of up to nearly 5.5 GeV/$c^2$ and identify events that are most probably magnetic particles up to a mass of 5.505 GeV/$c^2$.

## IV. SUMMARY

Given that Maxwell's equations can be made charge symmetric leads one to the conclusion that there might exist stable magnetic monopoles with magnetic charge of 1$e$. In fact, the model predicts a magnetic counterpart to all of the known electric fundamental particles. The two predicted stable magnetic monopole particles would then be 1) a counterpart to the electron (we call it the magneticon) and 2) a magnetically charged counterpart to the proton again with magnetic charge strength of 1$e$. We have not found any experiment that has ruled out the possible existence of either of these unit charge magnetic monopoles with the possible exception of the Crystal Ball detector while it was running on the $\Upsilon$(4S) at PETRA. The Crystal Ball collaboration did not see any enhancement in the muon production rate which is where we think a magneticon signal would have resided. We therefore have set a tentative lower limit on the mass of the lightest magnetic monopole (the magneticon) at about 4.5 GeV/$c^2$.

With this in mind, we encourage the BELLE II collaboration to perform a unique and explicit search for this 1$e$ magnetic charge possibility. We have modeled the trigger of the small cell section of the BELLE II drift chamber in a simple simulator and we have found that the BELLE II CDC can have a highly efficient trigger for this possible new particle essentially up to the kinematic limit of production or a mass of 5.505 GeV/$c^2$ for an $E_{cm}$ of 11.02 GeV. Even at this upper kinematic limit we predict that BELLE II should acquire 359 magneticon events for 10 fb$^{-1}$ of integrated luminosity (see Table 1). Of course, any higher $E_{cm}$ obtained by the accelerator would extend the search range for the magneticon.

## V. ACKNOWLEDGEMENTS

We would like to thank our colleagues for many fruitful discussions and observations as well as the BELLE II management members we have spoken to and who have shown an interest in this proposal.